\documentclass[useAMS,usenatbib]{mn2e}
\usepackage{epsfig}

\newcommand{\target}{4U\,1822--000}

\title[On the ultra-compact nature of 4U\,1822--000]
{On the ultra-compact nature of 4U\,1822--000}
\author[T. Shahbaz, C. A. Watson and Hernandez-Peralta]
{T.\,Shahbaz$^{1}$\thanks{E-mail: tsh@ll.iac.es}, 
C.A.\,Watson$^{2}$ and H.\,Hernandez-Peralta$^{1}$ \\
$^{1}$Instituto de Astrof\'\i{}sica de Canarias, 38200 La Laguna,
    Tenerife, Spain \\
$^{2}$Department of Physics and Astronomy, University of Sheffield,
    Sheffield, S3 7RH, UK \\
}

\begin{document}

\pagerange{\pageref{firstpage}--\pageref{lastpage}} \pubyear{2006}

\maketitle

\label{firstpage}

\begin{abstract}

\noindent
We report the discovery of a periodic modulation in the optical lightcurve of 
the candidate ultra-compact X-ray binary \target. Using time-resolved optical
photometry taken with the William Herschel Telescope we find evidence for a 
sinusoidal modulation with a semi-amplitude of 8\,percent and a period of
191\,min,  which is most likely close to the true  orbital period of the
binary.  Using the van Paradijs \& McClintock relation for the absolute
magnitude and the distance modulus allowing for interstellar  reddening,  we
estimate the distance to \target\ to be 6.3\,kpc. The long orbital period and 
casts severe doubts on the ultra-compact nature of \target.

\end{abstract}


\section{Introduction}

Low-mass X-ray binaries are systems in which a low-mass companion star
transfers material onto a neutron star or back hole. Most of the
systems have orbital periods of hours to days and contain ordinary
hydrogen-rich donor stars.  The minimum orbital period for systems
with hydrogen-rich donor stars  is around 80 min \citep{Rappaport82},
however, systems with hydrogen-poor or degenerate donor stars can
evolve to extremely small binary separations with orbital periods  as
short as a few minutes \citep{Nelson86}. Such systems are called
ultra-compact X-ray binaries (UCXBs) and have a range in orbital
periods from 11 to 50 minutes (see \citealt{NJ06})

X-ray spectroscopy has identified several candidate UCXBs based on 
similarities with the known UCXB 4U\,1850--87 where the inferred 
enhanced neon/oxygen ratio is interpreted as being local to the system. 
\citet{Juett01} thus concluded  that the candidate UCXBs have ultra-short 
periods and proposed that  their donors stars were originally 
carbon-oxygen or oxygen-neon-magnesium white dwarfs. There are 12 
systems with known or  suggested orbital periods, which include a transient 
pulsar, X-ray busters  and millisecond pulsars  UCXB candidates have also 
been identified via their low optical-to-X-ray flux ratio and optical 
faintness. This is what is expected for the small  accretion discs in these 
systems, because the optical emission is dominated by the irradiated accretion 
disc whose surface is determined by the size of the binary i.e. the orbital 
period (for an observational review see  \citealt{NJ06}).

\target\ was discovered by Uhuru and has since been observed briefly  
by all the major X-ray missions  (\citealt{Giacconi72};
\citealt{Warwick81}; \citealt{Wood84}; \citealt{SL92};  \citealt{CS97}) and
has been roughly constant  in the X-rays with an unabsorbed flux of $\rm
1\times\,10^{-9}\,ergs\,cm^{-2}\,s^{-1}$  in the 0.5--20\,keV range
\citep{CS97}.   It has a faint $V$=22 optical counterpart \citep{CI85} and was
identified as an UCXB candidate based on its  low X-ray/optical flux ratio.
Chandra observations show that the X-ray source is coincident with the optical
position and there no evidence for orbital modulation in the X-rays
(\citealt{Juett05}). \citet{Wang04} obtained optical photometry of the
optical counterpart  using the 6.5m Magellan telescope and  detected 
significant variability on a  timescale of about 90\,mins.  In this letter we
report on our optical  time-resolved photometry of \target.

%
\begin{figure*}
\psfig{angle=-0,width=15.0cm,file=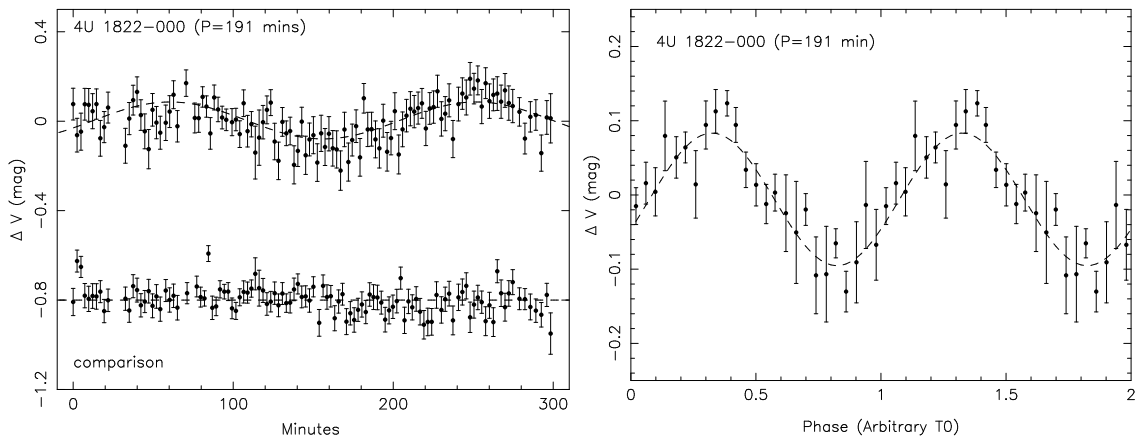}
\caption{ Left: the lightcurve of \target. The dashed line is a
sinusoidal fit with a period of 191\,min. The comparison star is also
shown and is shifted vertically downwards for clarity. The lightcurves
of both \target\ and the comparison star have been binned to a time
resolution of 2.4\,min.  The dashed line is a sinusoidal and constant
fit to the \target\ and comparison star lightcurves
respectively. Right: the lightcurve folded on the 191\,min period and
binned into 25 phase bins. The dashed line is a sinusoidal fit to the
data and two orbital cycles are shown.}
\label{FIG:LCURVE}
\end{figure*}

\section{Observations and Data Reduction}
\label{OBSERVATION}

Our optical photometric observations were taken on 2006 July 23 using
the 4.5m William Herschel Telescope on La Palma. We used the TEK
1024$\times$1024 pixel CCD camera at aux-port focus, 
A CCD binning of 2$\times$2 was used providing
a plate scale of 0.22 arcsec\,pixel$^{-1}$ and an unvignetted (circular) field
diameter of 1.8\,arcmins. The Johnson $V$-band filter was used and the total
observation  time length  on \target\ was approximately 5\,h. The exposure time
of each individual frame was 60\,s and with a readout time of 13\,s, we
obtained one image approximately every 1.2\,min over the course of our
observations.  The conditions were not photometric because of the calima
(Sahara dust), but the seeing  was excellent, ranging from 0.6 to 1.0\,arcsecs.

We used IRAF  for our initial data reduction, which included bias subtraction
using the overscan regions of the CCD, and flat-fielding using sky flat fields
taken during twilight. The ULTRACAM reduction  pipeline software was then used
to obtain lightcurves for \target\ and several comparison stars by extracting
the counts using aperture photometry. A variable aperture which scaled with the
seeing was used.  Differential lightcurves were then obtained by computing the 
count ratio of \target\ with respect to a local standard (non-variable star).
As a check of the photometry and systematics in the reduction procedure, we
also extracted lightcurves of a comparison star  similar in brightness to the
target.   The photometric accuracy of \target\ for each exposure is
about 10 percent and agrees with the scatter of the comparison star with
similar brightness.

\section{Results}
\label{RESULTS}

Figure\,1 shows the lightcurve for \target\ which clearly  exhibits a
strong modulation. Given the baseline of our dataset of 5\,h, a Lomb-Scargle
analysis of the lightcurve does not reveal any significant periods on
timescales of less than a few hours. However, fitting the lightcurve with a
sinusoidal modulation reveals a period of 191$\pm$8\,min and a semi-amplitude
of 0.083$\pm0.010$\,mags.  Although the determination of the
period  is biased because our baseline is just less than twice the period, it
should be noted that \citet{Wang04} also saw a modulation on a timescale of
$\sim$180\,min  (2$\times$90\,min; note that they observed half a periodic
modulation). 

\section{Discussion}

We have found a 191\,min sinusoidal modulation  in the  optical
lightcurve of \target\ which most likely arises from either X-ray
irradiation of the inner face of the secondary star and/or a superhump
modulation from the accretion disc. Sumperhumps only occur in binaries
with extreme mass ratios  (such as UCXBs), and are due to a
precessing, eccentric accretion disk  \citep{WK91}, which have
observed optical periods a few percent longer than the binary orbital
period.  Thus the 191\,min period most likely lies close to the
orbital period of the binary, and we assume this to be the case in the
rest of the discussion.

\citet{VM94} determined the absolute magnitude of X-ray binaries and derived an
empirical relation between the absolute magnitude ($M_{\rm V}$),  orbital
period ($P$) and X-ray luminosity ($L_{\rm X}$)

\begin{equation}
M_{\rm V} = 1.57 - 2.27\log(\Sigma); \\
\Sigma = (L_{\rm X}/ L_{\rm Edd})^{1/2} (P/1hr)^{2/3}
\end{equation}

\noindent 
where  $L_{\rm Edd}$ is the Eddingtion luminosity  
(2$\times 10^{38}$\,erg\,s$^{-1}$ for a 1.4\,$\rm M_{\odot}$ neutron star).  
Their relation was based on the assumption that the optical flux is due to
X-ray irradiation of the accretion disc, which means that UCXBs with short
orbital periods are expected to have relatively faint discs. Therefore faint 
$\rm M_{\rm V}$ values have been used to select UCXB candidates  which have 
$\rm M_{\rm V}>$3.7 \citep{NJ06}. Indeed \target\ was provisionally classified
as an UCXB based on its' faint  absolute magnitude; $\rm M_{\rm V}>$7.5
assuming a distance of 8\,kpc   (even if the distance is 20\,kpc,  
$\rm M_{\rm V}$=5.5) and the lack of hydrogen and helium emission lines in it's
low quality VLT optical spectrum \citep{Nelemans06}. However, our estimate of
the orbital period of 191\,min casts severe doubts on the UCXB nature of
\target.

Based on our estimate for the orbital period and  assuming that \target\ is a
persistent X-ray binary,  we can use  the van Paradijs \& McClintock relation
and the distance modulus to  estimate the distance to \target.  The optical
magnitude of \target\ is $V$=22 \citep{CI85} and the optical extinction can be
determined using the empirical relation between the optical extinction 
($\rm A_{\rm V}$) and the total column density of hydrogen $\rm N_{\rm H}$ 
(\citealt{G75}; \citealt{PS95}). Using the observed value for 
$\rm N_{\rm H}=0.97\times 10^{22}\,cm^{-2}$ \citep{Juett05} we find  
$\rm A_{\rm V}$=5.5\,mag. Using the de-reddended optical magnitudes,  
the unabsorbed X-ray flux of 1.1$\rm \times 10^{-9}\,ergs\,cm^{-2}\,s^{-1}$
(determined using the observed  X-ray flux, the power-law index and 
$\rm N_{\rm H}$ given in  \citealt{Juett05}) and $P$=191\,min in conjuction
with equation (1) and the distance modulus,  we find the distance to \target\
to be 6.3$\pm$2.0\,kpc and  $\rm M_{\rm V}=$2.6$^{+0.8}_{-0.6}$ 
(the  errors relfect the uncertainties in the van Paradijs \&
McClintock  relation). 

We can use the distance and the X-ray flux to estimate the observed  mass
accretion rate, which can be compared to theoretical mass transfer rate
expected for gravitational radiation and magnetic braking. The
unabsorbed X-ray flux and the distance of 6.3\,kpc give the accretion
luminosity which  suggests an observed  mass accretion rate of 
$\rm \dot{M}_{\rm obs} = 9^{+8}_{-5.0} \times (\eta/0.1)^{-1}$(d/6.3\,kpc)$^2 
\times  10^{-10}\,M_{\odot}\,yr^{-1}$, where $\eta\sim0.1$ (=$GMc^2/R$) is the
efficiency  of converting accretion into radiation for a neutron star. This is
comparable with estimates for the mass transfer rate onto a   1.4\,$\rm
M_{\odot}$ neutron star  driven by a combination of  magnetic braking and 
gravitational  radiation for a main sequence star \citep{KKB96}.

Finally, assuming that the 191\,min period is the orbital period then we can
estimate the mean density (and spectral type) of a Roche-lobe filling
main-sequence in  such a binary;  $\rho/\rho_{\odot}= 7.75P_{\rm hr}^{-2}$
\citep{Smith98}. For \target\  we find $\rho$=7.5$\rho_{\odot}$ which
corresponds to a M4--5 dwarf  star \citep{Allen76}.

\section{Is \target\ an UCXB?}

The Chandra HEG and MEG X-ray spectrum of \target\ can be fitted with a
power-law and blackbody model and does not show any features 
that may indicate that it is an UCXB. Furthermore, our
determination of a 191\,min period (which we believe is close to the orbital
period)  casts severe doubts on the UCXB nature of \target. Given this, it is
puzzling that no hydrogen or helium emission lines are seen in the optical VLT
spectra. However, this could be due to the poor quality of the  spectra
\citep{Nelemans06}. 

It is thus interesting to make a comparison with the similar short
period XRBs,  such as 4U\,1323-619, EXO\,0748--676 and GR\,MUS
\cite{Ritter03},  with orbital periods of 2.96\,hr, 3.82\,hr and 3.93\,hr
respectively.  The optical lightcurves of these systems, which are all X-ray
bursters,  show a sinusoidal modulation with partial eclipes/dips consistent
with the  idea that a large reprocessing region in the disk is being partially
obscured by structure at  the disk edge. In all these system, broad emission
lines are observed. In contrast the optical lightcurve of \target\ contains
only a sinusoidal component,  which suggests that system lies at a relatively
low inclination angle. However, this  does not explain the lack of emission
lines, which one would expect to be independent  on orbital phase for systems
at low inclination angles.  Clearly a more detailed photmetric and
spectroscopic study will prove useful.

\section*{Acknowledgments}

TS acknowledges support from the Spanish Ministry of Science  and
Technology  under the programme Ram\'{o}n y Cajal and 
under the project AYA2004-02646. CAW is supported
by a PPARC postdoctoral fellowship.

\end{document}